\documentclass[fleqn,twoside]{article}
\usepackage{espcrc2}

\usepackage{graphicx}
\usepackage[figuresright]{rotating}

\newcommand{\AmS}{{\protect\the\textfont2
  A\kern-.1667em\lower.5ex\hbox{M}\kern-.125emS}}

\title{The role of the water molecules in novel superconductor, Na$_{0.35}$CoO$_{2}\cdot$1.3H$_{2}$O}

\author{H. Sakurai\address[SMC]{Superconducting Materials Center, National Institute for Materials Science, 1-1 Namiki, Tsukuba, Ibaraki, 305-0044, Japan.}%
        \thanks{SAKURAI.Hiroya@nims.go.jp},
        K. Takada\address[AML]{Advanced Materials Laboratory, National Institute for Materials Science, 1-1 Namiki, Tsukuba, Ibaraki, 305-0044, Japan.},
		F. Izumi\addressmark[AML],
		D. A. Dilanian\addressmark[AML],
		T. Sasaki\addressmark[AML],
        and
        E. Takayama-Muromachi\addressmark[SMC]
}
       
\begin{document}

\begin{abstract}
In order to investigate the role of the water molecules in Na$_{0.35}$CoO$_{2}\cdot$1.3H$_{2}$O, we synthesized superconducting Na$_{0.35}$CoO$_{2}\cdot$1.3H$_{2}$O and nonsuperconducting Na$_{0.35}$CoO$_{2}\cdot$0.7H$_{2}$O, and measured their normal-state magnetic susceptibilities. The susceptibility of Na$_{0.35}$CoO$_{2}\cdot$1.3H$_{2}$O has an enhancement below ~150 K probably caused by ferromagnetic fluctuation, whereas no such enhancement was observed in Na$_{0.35}$CoO$_{2}\cdot$0.7H$_{2}$O. The water molecules in Na$_{0.35}$CoO$_{2}\cdot$1.3H$_{2}$O may work to shield random coulomb potential of the Na ions with smoother potential at the CoO$_{2}$ layer. This effect may account for the appearance of superconductivity in Na$_{0.35}$CoO$_{2}\cdot$1.3H$_{2}$O.
\vspace{1pc}
\end{abstract}

\maketitle

There exist a large group of layered compounds with the chemical formula of A$_{x}$MO$_{2}$ ($0\leq x\leq1$) (A: Alkali metal, Cu, Ag, and so forth, M: transition metal, Al, Ga, and so forth). In these compounds, M atom is octahedrally coordinated by six oxygen atoms, and the octahedra are connected by each other sharing edge to compose a two-dimensional MO$_{2}$ layer. The M sites of the MO$_{2}$ layer form a triangular lattice. The coordination of the A site varies depending on the structural type; for example, octahedral coordination in an $\alpha$-NaFeO$_{2}$-type structure, dumbbell coordination in a delafossite structure, etc.

In many cases, the M ions in A$_{x}$MO$_{2}$ has a mixed-valent state as expected from the composition. Due to the layered structures and the mixed valent states, this family of compounds show a variety of physical and chemical properties. LiCoO$_{2}$ is an electrode material practically employed for rechargeable battery due to its high mobility of Li ions which is caused by the layered structure. K$_{0.45}$MnO$_{2}$ is well known because MnO$_{2}$ nano-sheets have been derived from it \cite{MnNano}. LiVO$_{2}$ is considered to be a compound which shows orbital order \cite{LiVO2} and CuFeO2 shows a complex magnetic phase diagram \cite{CuFeO2} with both properties originated from the triangular lattices of the M atoms.

Na$_{x}$CoO$_{2}$ is one of the most attractive compounds in the family. In this system, Co is between trivalent ($3d^{6}$) and tetravalent ($3d^{5}$), and low-spin configuration is expected from the Co-O distance \cite{CoOdist}. Thus, the CoO$_{2}$ layer can be regarded as an electron-doped $S=1/2$ triangular lattice.  The compound with $x=0.5$ is famous for the large thermoelecric power \cite{NaCo2O4}. It has a relatively large magnetic susceptibility of about $3\times10^{-3}$ emu/Co mol with weak temperature dependence. Moreover, spin-density-wave (SDW) like transition has been observed after 20\% 
substitution of Cu for Co. These facts indicate that this compound is a strongly correlated metal. Similar properties have been observed for compounds with other x values. In case of $x=0.7$, charge ordered state has been observed \cite{x07}. The magnetic susceptibility of $x=0.75$ has Curie-Weiss-like term, whose Curie constant cannot be explained by local moments, and shows SDW-like transition at 22 K \cite{x075}.

\begin{figure*}[htb]
\includegraphics[width=14cm,keepaspectratio]{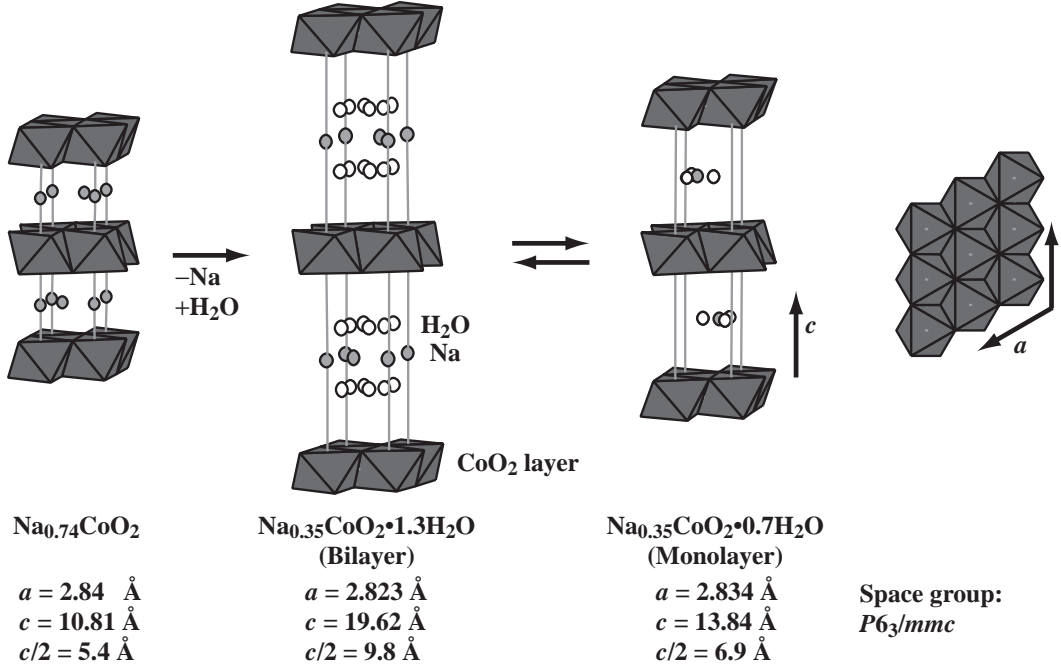}
\caption{
Structure of Na$_{0.7}$CoO$_{2}$, Na$_{0.35}$CoO$_{2}\cdot$1.3H$_{2}$O (NCO1.3H2O), and Na$_{0.35}$CoO$_{2}\cdot$0.7H$_{2}$O (NCO0.7H2O). The dark octahedron, the gray circle, and the open circle represent CoO6, Na, and H2O, respectively.
}
\label{struct}
\end{figure*}

Recently, Na$_{0.35}$CoO$_{2}\cdot$1.3H$_{2}$O (NCO1.3H2O) was synthesized and it was found to undergo superconducting transition at $T_{C} \sim 4.6$ K \cite{NCO}. The two-dimensional nature of structure is far more pronounced in this compound where two adjacent CoO$_{2}$ layers are separated by thick insulating block composed of three layers of Na and H$_{2}$O with the distance of about 0.98 nm (see Fig. \ref{struct}). Its superconductivity looks to be unconventional judged from unusual phenomenological superconducting parameters \cite{NCO}. Moreover, no Hebel-Schlichter peak (coherence peak) has been observed in the temperature dependence of spin-lattice relaxation time of nuclear quadrupole resonance (NQR) \cite{NQR}. This suggests that the superconducting gap has line-nodes denying simple s-wave symmetry. Concerning whether Knight shift changes at $T_{C}$ or not, there are contradictious reports \cite{KightShift} and it is still unclear whether the Cooper pair is spin-singlet or triplet. On the other hand, theoretical studies have suggested various possibilities on the symmetry of superconductivity: $p+ip'$-, $d+id'$-, and $f$- waves \cite{theory}. 

Quite recently, a new analogue was discovered related to NCO1.3H2O;  it is Na$_{0.35}$CoO$_{2}\cdot$0.7H$_{2}$O (NCO0.7H2O) \cite{NCO07H2O} where the CoO$_{2}$ layers are separated by a single plane composed of H$_{2}$O and Na rather than the triple planes in NCO1.3H2O (see Fig. \ref{struct}). One may suppose that NCO0.7H2O is also superconducting because the CoO$_{2}$ layers are still separated largely with a distance of 0.69 nm, which is larger than the distance between CuO$_{2}$ planes of La$_{2}$CuO$_{4}$ (0.66 nm) \cite{La214}. However, superconductivity is completely suppressed in NCO0.7H2O and it is implied that water molecules in NCO1.3H2O plays a certain important role for superconductivity besides the separation of the CoO$_{2}$ layers. We report here normal-state magnetic susceptibilities of NCO1.3H2O and NCO0.7H2O and discuss their difference in connection with the role of the water molecules.
\begin{table*}[htp]
\caption{
Parameters obtained by the Curie-Weiss fit.
}
\begin{tabular}{||c|c|c||} \hline
	& NCO1.3H2O & NCO0.7H2O \\ \hline
$\chi_{0}$ ($10^{-4}$ emu/mol) & 2.95 & 4.31 \\ \hline
$A$ ($10^{-10}$ emu/mol K$^{2}$) & 9.37 & 9.36 \\ \hline
$C$ ($10^{-3}$ emu K/mol) & 8.36 & 1.93 \\ \hline
$\theta$ (K) & $-24.3$ & $-0.00459$ \\ \hline
\end{tabular}
\label{parameters}
\end{table*}

The powder samples of NCO1.3H2O and NCO0.7H2O were synthesized through a soft chemical process \cite{NCO,NCO07H2O}. A mother oxide Na$_{0.7}$CoO$_{2}$ was prepared from Na$_{2}$CO$_{3}$ (99.99\%) 
and Co$_{3}$O$_{4}$ (99.9\%) 
by solid-state reaction. A well-pulverized powder of Na$_{0.7}$CoO$_{2}$ was immersed in the Br$_{2}$/CH$_{3}$CN solution for 5 days to deintercalate Na$^{+}$ ions, then the product was filtered, washed with CH$_{3}$CN and distilled water, and finally dried in 70\%
humidity atmosphere. The obtained samples were identified as a single phase of NCO1.3H2O. The NCO0.7H2O sample was obtained by storing NCO1.3H2O under a N$_{2}$ gas flow for 6 days. Qualities of the samples thus prepared were checked by x-ray powder diffraction.
\begin{figure}[htp]
\includegraphics[width=7cm,keepaspectratio]{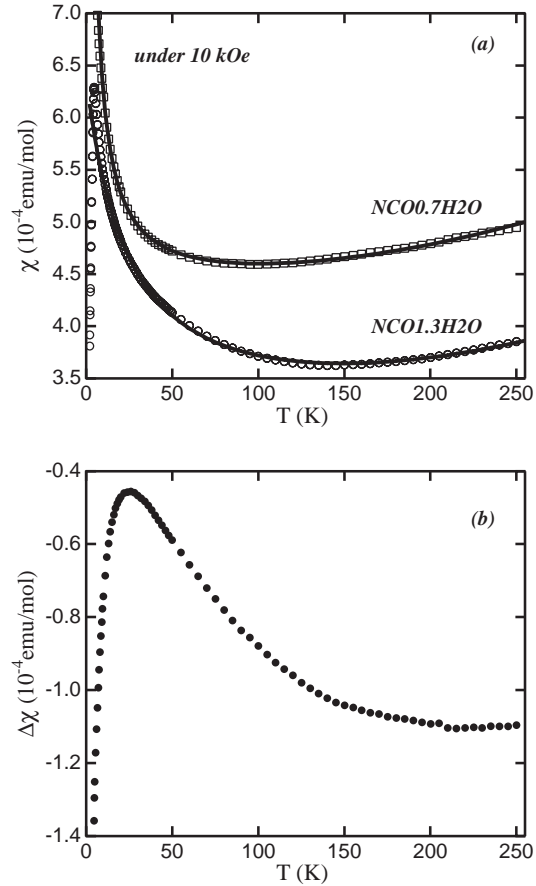}
\caption{
(a) Magnetic susceptiblilties of Na$_{0.35}$CoO$_{2}\cdot$1.3H$_{2}$O (NCO1.3H2O) and Na$_{0.35}$CoO$_{2}\cdot$0.7H$_{2}$O (NCO0.7H2O). (b) the difference between the magnetic susceptibilies of NCO1.3H2O and NCO0.7H2O (the former minus the latter).
}
\label{chi}
\end{figure}

The magnetic susceptibility curves of NCO1.3H2O and NCO0.7H2O are shown in Fig. \ref{chi}(a); they have a broad minimum around 130 K and 95 K, respectively. At first, we assumed simply Pauli paramagnetic term and Curie-Weiss term in which the latter was caused by magnetic impurities and/or lattice defects. Thus, the data between 10 K and 250 K were fit by the equation: $\chi (T) = \chi _{0}+AT^{2}+C/(T-\theta)$. The obtained parameters for NCO1.3H2O and NCO0.7H2O are shown in table \ref{parameters}. The difference between the susceptibilities, $\Delta\chi = \chi_{NCO0.7H2O}-\chi_{NCO1.3H2O}$, are shown in Fig. \ref{chi}(b).

From the table \ref{parameters} and Fig. \ref{chi} (b), it is seen that Pauli paramagnetic terms are not so different with each other, whereas Curie-Weiss terms are largely different. The variation of Weiss temperature is worth noting, because it is related to bulk natures, such as the interaction between local moments and spin fluctuation of itinerant electron system. The NCO0.7H2O phase was formed just by drying the NCO1.3H2O phase without any high temperature treatment. Weiss temperature will not vary by water molecules adsorbed on the surface of the grain. It will vary only when water molecules are intercalated (deintercalated) into (from) the bulk. Thus, it is reasonable to conclude that the Curie-Weiss terms are not fully originated form magnetic impurity phase(s) but they reflect at least partly some intrinsic natures. In particular, the Curie-Weiss-like enhancement below $\sim$150 K in NCO1.3H2O, which disappears in NCO0.7H2O (see Fig. \ref{chi}(b)), should have intrinsic origin. Microscopic measurements such as NQR and $\mu$SR support this conclusion \cite{NQR,KightShift}.

The enhancement below $\sim$150 K is considered to be due to ferromagnetic spin fluctuation, because only ferromagnetic component can be observed by the static measurement. Tendency of ferromagnetic order in this compound has been already pointed out by Singh \cite{BandCal}. Water molecules in NCO1.3H2O look to reinforce the ferromagnetic fluctuation. Similar enhancement of the magnetic susceptibility is seen in La-doped Sr$_{2}$RuO$_{4}$ \cite{Sr214} and Na$_{0.5}$CoO$_{2}$ \cite{Na05CoO2}.

As seen in Fig. \ref{chi}(b), Dc decreases below 25 K, which may suggest that the enhancement of c in NCO0.7H2O is intrinsic and is also caused by the ferromagnetic fluctuation. If the enhancement is due to a magnetic impurity phase(s), the decrease of $\Delta\chi$ can not be explained because the impurity should exist in NCO1.3H2O, as well. Preliminary $\mu$SR measurements of NCO0.7H2O suggest that this enhancement has indeed intrinsic nature \cite{muSR}. In NCO0.7H2O, superconductivity may be observed at extremely low temperature.
\begin{figure}[t]
\includegraphics[width=7cm, ,keepaspectratio]{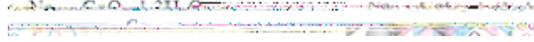}
\caption{
Isosurfaces of electron distribution of Na$_{0.35}$CoO$_{2}\cdot$1.3H$_{2}$O (NCO1.3H2O) and Na$_{0.35}$CoO$_{2}\cdot$0.7H$_{2}$O (NCO0.7H2O) seen from 110 direction [(a), (c), respectively]. Contour lines of electron distribution of NCO1.3H2O and NCO0.7H2O cut at (2/3,0,0) [(b), (c), respectively].
}
\label{MEM}
\end{figure}

Figure \ref{MEM} displays the electron densities of NCO1.3H2O and NCO0.7H2O obtained by maximum entropy method (MEM) on powder x-ray diffraction data \cite{NCO,NCO07H2O}. As seen in the figure, the electrons of water molecules seem to be attracted by the Na ions in both structures. This phenomenon is related to the polarization of the water molecule. In case of NCO1.3H2O, the water molecule is placed between the Na plane and the CoO$_{2}$ layer, and the coulomb potential of the Na ions may be shielded by the polarized water molecules to yield less pronounced random potential at the CoO$_{2}$ layer. This is very different from the situation of NCO0.7H2O where the water molecules and the Na ions are placed on the same plane. In general, physical properties of a two-dimensional electron system would be affected seriously by random potential. The disappearance of spin fluctuation in NCO0.7H2O may be caused by the random potential and the disappearance of superconductivity may be also caused by it.

In summary, we synthesized the powder samples of superconducting NCO1.3H2O and nonsuperconducting NCO0.7H2O, and measured their normal-state magnetic susceptibilities. Enhancement of the susceptibility below $\sim$150 K due to ferromagnetic spin fluctuation was observed in NCO1.3H2O but not in NCO0.7H2O. From the MEM analyses, it is suggested that the water molecules in NCO1.3H2O work to shield the random coulomb potential of the Na ions, providing smoother potential at the CoO$_{2}$ layer. This effect of water molecules may account for the appearance of the ferromangetic fluctuation and superconductivity in NCO1.3H2O.

We thank A. Tanaka, T. Waki, C. Michioka, M. Kato, K. Yoshimura, K. Ishida, and W. Higemoto for their valuable suggestions on this study. This work has been partially supported by CREST of JST (Japan Science and Technology Agency). One of authers (H. S.) has been supported by Japan Society for the Promotion of Science.

\end{document}